\documentclass[prl,twocolumn,showpacs,preprintnumbers,amsmath,amssymb]{revtex4}

\usepackage{graphicx}
\usepackage{dcolumn}
\usepackage{bm}

\begin{document}

\newcommand*{\cm}{cm$^{-1}$\,}
\newcommand*{\Tc}{T$_c$\,}


\title{Superconductivity in hole-doped (Sr$_{1-x}$K$_x$)Fe$_2$As$_2$}
\author{G. F. Chen}
\author{Z. Li}
\author{G. Li}
\author{W. Z. Hu}
\author{J. Dong}
\author{X. D. Zhang}
\author{P. Zheng}
\author{N. L. Wang}
\author{J. L. Luo}

\affiliation{Beijing National Laboratory for Condensed Matter
Physics, Institute of Physics, Chinese Academy of Sciences,
Beijing 100190, China}


\begin{abstract}

A series of layered (Sr$_{1-x}$K$_x$)Fe$_2$As$_2$ compounds with
nominal x=0 to 0.40 are synthesized by solid state reaction
method. Similar to other parent compounds of iron-based pnictide
superconductors, the pure SrFe$_2$As$_2$ shows a strong
resistivity anomaly near 210 K, which was ascribed to the
spin-density-wave instability. The anomaly temperature is much
higher than those observed in LaOFeAs and BaFe$_2$As$_2$, the two
prototype parent compounds with ZrCuSiAs- and ThCr$_2$Si$_2$-type
structures. K-doping strongly suppresses this anomaly and induces
superconductivity. Like in the case of K-doped BaFe$_2$As$_2$,
sharp superconducting transitions at T$_c\sim$38 K was observed.
We performed the Hall coefficient measurement, and confirmed that
the dominant carriers are hole-type. The carrier density is
enhanced by a factor of 3 in comparison to F-doped LaOFeAs
superconductor.
\end{abstract}



\maketitle

The recent discovery of superconductivity with transition
temperature T$_c\sim$26 K in LaO$_{1-x}$F$_x$FeAs has generated
tremendous interest in scientific community\cite{Kamihara08}.
Shortly after this discovery, the \Tc was raised to 41-55 K by
replacing La by rare-earth Ce, Sm, Pr, Nd, etc, making those
systems the first non-copper based materials with T$_c$ exceeding
50 K.\cite{Chen1,XHChen,Ren1,Ren2,CWang} The undoped quaternary
compounds crystallize in a tetragonal ZrCusiAs-type structure,
which consists of alternate stacking of edge-sharing Fe$_2$As$_2$
tetrahedral layers and La$_2$O$_2$ tetrahedral layers along
c-axis. Except for a relatively high transition temperature, the
system displays many interesting properties. Most remarkably, the
superconductivity was found to be in the vicinity of a
spin-density-wave (SDW) instability.\cite{Dong,Cruz} The
competition between superconductivity and SDW instability was
first identified in LaO$_{1-x}$F$_x$FeAs system\cite{Dong}, and
subsequently found in other rare-earth substituted
systems.\cite{Chen1,Ding,Chen2} Those observations reveal that the
superconductivity was intimately related to the magnetic
fluctuations.

Very recently, it is found that the ternary iron arsenide
BaFe$_2$As$_2$ with a tetragonal ThCr$_2$Si$_2$-type structure,
which contains identical edge-sharing Fe$_2$As$_2$ tetrahedral
layers as LaOFeAs, exhibits a similar SDW instability at 140 K,
which was characterized by strong anomalies in resistivity,
specific heat, magnetic susceptibility, and structural
distortion.\cite{Rotter1} It is therefore suggested that
BaFe$_2$As$_2$ could serve as a new parent compound for ternary
iron arsenide superconductors. Indeed, soon after that, the
superconductivity with T$_c\sim$38 K was found in K-doped
BaFe$_2$As$_2$, which was suggested to be a hole-doped iron
arsenide superconductor, although a direct measurement of Hall
coefficient is lacking.\cite{Rotter2} Before K-doped
BaFe$_2$As$_2$, the superconductivity induced by hole-doping was
reported in (La$_{1-x}$Sr$_x$)FeAs system.\cite{Wen}

In this work, we report fabrication of a series of layered
(Sr$_{1-x}$K$_x$)Fe$_2$As$_2$ compounds with nominal x=0, 0.1,
0.2, and 0.40 by solid state reaction method. We expect to see
similar phenomena as observed in (Ba$_{1-x}$K$_x$)Fe$_2$As$_2$
compounds. Indeed, the pure SrFe$_2$As$_2$ shows a strong SDW
anomaly in resistivity. However, the anomaly occurs near 210 K,
being substantially higher than that observed on LaOFeAs and
BaFe$_2$As$_2$, the two prototype parent compounds with ZrCuSiAs-
and ThCr$_2$Si$_2$-type structures. K-doping strongly weakens this
anomaly and induces superconductivity. We performed the Hall
coefficient measurement on K-doped superconducting sample, and
confirmed that the dominant carriers are hole-type. Furthermore,
the carrier density is much higher than F-doped
LaO$_{0.9}$F$_{0.1}$FeAs superconductor.

The polycrystalline samples were synthesized by solid state
reaction method using SrAs, KAs, Fe$_2$As as starting materials.
The synthesizing method is similar to that described in our
earlier paper\cite{Chen1}. SrAs was prepared starting from
reaction involving Sr chips and As pieces at 500 $^{\circ}C$ for
10 hours and then 750 $^{\circ}C$ for 20 hours. KAs was prepared
by reacting K lumps and As pieces at 600 $^{\circ}C$ for 24 hours.
In order to avoid Sr and K attack on the quartz tubes at elevated
temperatures, the elements were put into alumina crucibles and
finally sealed in quartz tubes under Ar gas atmosphere. The
obtained materials SrAs, KAs, Fe$_2$As were thoroughly mixed in a
correct ratio and pressed into pellets. The pellets were wrapped
with Ta foil and sealed in quartz tube under Ar gas atmosphere.
They were then annealed at 750-900 $^{\circ}C$ for 24 hours. The
resulting samples were characterized by a powder X-ray
diffraction(XRD) method with Cu K$\alpha$ radiation at room
temperature. The electrical resistivity was measured by a standard
4-probe method and the Hall coefficient was measured by a 5-probe
method. The ac magnetic susceptibility was measured with a
modulation field in the amplitude of 10 Oe and a frequency of 333
Hz. These measurements were preformed in a Physical Property
Measurement System(PPMS) of Quantum Design company.

Figure 1 shows the XRD patterns for the parent compound. The
diffraction peaks could be well indexed on the basis of tetragonal
ThCr$_2$Si$_2$-type structure with the space grounp I4/mmm. A
little impurity phases detected from the measurements were
attributed to the unstable behavior of SrFe$_2$As$_2$ in air. The
lattice constants were a = 0.3920 nm and c = 1.240 nm, consistent
with the reported values\cite{Pfis}.

\begin{figure}[b]
\includegraphics[width=8cm,clip]{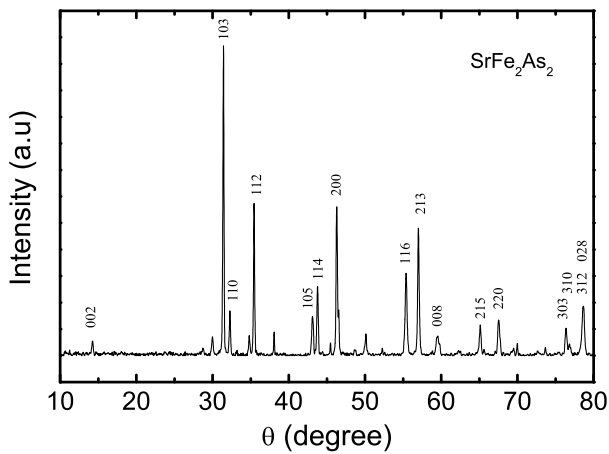}
\caption{(Color online) X-ray powder diffraction patterns for
SrFe$_2$As$_2$ sample.}
\end{figure}

Figure 2(a) shows the temperature dependence of the resistivity
for (Sr$_{1-x}$K$_x$)Fe$_2$As$_2$. The pure SrFe$_2$As$_2$ sample
exhibits a strong anomaly near 210 K: the resistivity drops
steeply below this temperature. This is a characteristic feature
related to SDW instability.\cite{Dong} We found that the anomaly
occurs at substantially higher temperature than those seen on
LaOFeAs and BaFe$_2$As$_2$. Krellner et al. also found similar
anomaly very recently, and addressed its relation to the SDW-type
magnetic ordering.\cite{Krellner} The superconductivity with
T$_c\sim$38 K was observed for K-doped samples with x=0.2 and 0.4.
The residual anomaly at high temperature comes from the
inhomogeneity of samples. It can be removed by further annealing
process (powered and pressed again), but this treatment decreases
the superconducting transition temperature. Figure 2(b) shows the
temperature dependence of resistivity for the reannealed sample
Sr$_{0.6}$K$_{0.4}$Fe$_2$As$_2$. The T$_c$ is depressed from 38 K
to 20 K. The bulk superconductivity in K-doped SrFe$_2$As$_2$ is
confirmed by ac magnetic susceptibility measurements. Figure 2(c)
shows the real and imaginary parts of ac susceptibility in a
temperature range near T$_c$ for the reannealed sample
Sr$_{0.6}$K$_{0.4}$Fe$_2$As$_2$.

\begin{figure}[t]
\centerline{\includegraphics[width=3.2in]{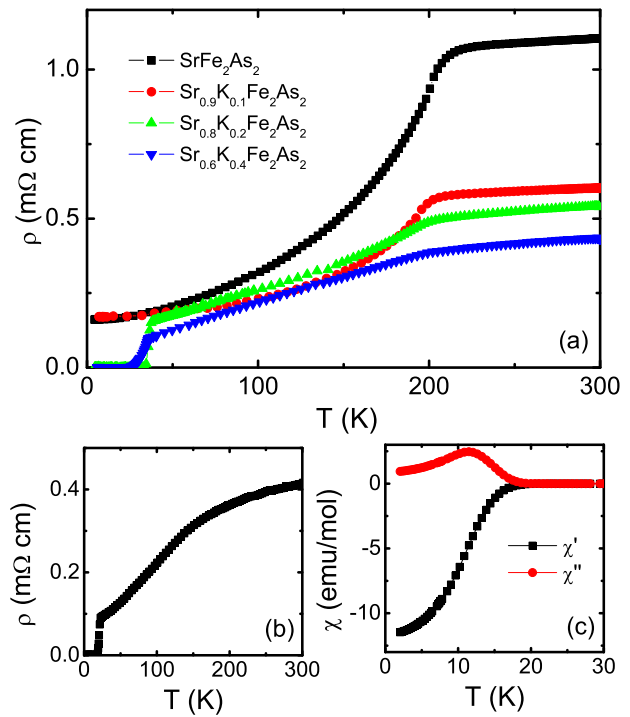}}%
\caption{(Color online) (a) The electrical resistivity vs
temperature for a series of (Sr$_{1-x}$K$_x$)Fe$_2$As$_2$. (b) The
electrical resistivity vs temperature for the reannealed
Sr$_{0.6}$K$_{0.4}$Fe$_2$As$_2$. (c) Real and imaginary parts of
T-dependent ac magnetic susceptibility for the reannealed
Sr$_{0.6}$K$_{0.4}$Fe$_2$As$_2$.}
\end{figure}

In (Sr$_{1-x}$K$_x$)Fe$_2$As$_2$, the replacement of Sr$^{2+}$
with K$^+$ adds extra holes to the system, the conducting carriers
are expected to be hole-type. To confirm this, we preformed Hall
coefficient measurement in the normal state. Figure 3 shows the
Hall coefficient versus T between 30 and 200 K of the reannealed
Sr$_{0.6}$K$_{0.4}$Fe$_2$As$_2$. The inset shows Hall voltage
driven by magnetic field at 100K in field up to 5T. A linear
dependence of the transverse voltage against applied magnetic
field is observed. The experiment indicates that the Hall
coefficient is positive in the measured temperature, confirming
the hole-type conducting carriers for
(Sr$_{1-x}$K$_x$)Fe$_2$As$_2$. This is strong contrast to that of
La(O$_{1-x}$F$_x)$FeAs where the conducting carriers are found to
be electron-type.\cite{Chen3}

\begin{figure}
\includegraphics[width=3.2in]{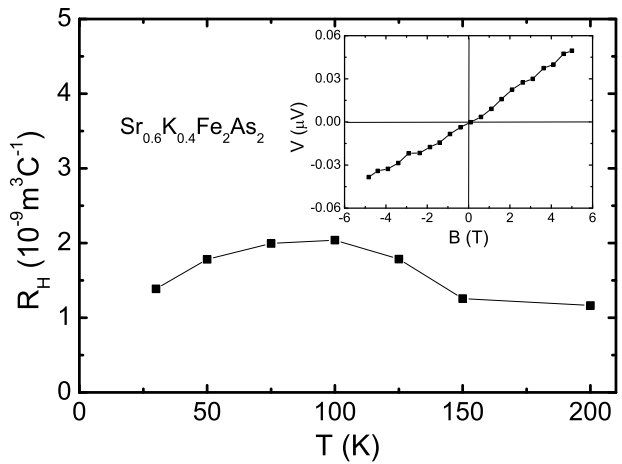}
\caption{(Color online) Hall coefficient R$_H$ vs temperature for
the reannealed Sr$_{0.6}$K$_{0.4}$Fe$_2$As$_2$. The inset shows
that the transverse voltage measured at T=100 K is proportional to
the applied magnetic field.}
\end{figure}

It is interesting that the Hall coefficient shows
temperature-dependent with a broad peak at about 100K. It is
possible that the T-dependence of R$_H$ comes from the multiple
bands effect. In LaOFeAs, the band calculations show that all the
five Fe d-orbital energy levels are not fully occupied, they cross
the Fermi level E$_F$, leading to five Fermi
surfaces.\cite{Leb07,Singh,fang} Here for
(Sr$_{1-x}$K$_x)$Fe$_2$As$_2$, we would expect to have similar
five Fermi surfaces. When the carrier scattering rates change with
temperature at different rates for different bands, R$_H$ can
become T-dependent. The carrier density is estimated to be
5.4$\times$10$^{21}$/cm$^3$ at 200K if we simply adopt a
single-band formula n=1/R$_H$e. The carrier density is more than 3
times as large as that of electron-doped single FeAs layer
LaO$_{0.9}$F$_{0.1}$FeAs with n$\sim$ 1.8$\times$10$^{21}$/cm$^3$
\cite{Chen3} at 200K.

To conclude, we have synthesized a series of layered
(Sr$_{1-x}$K$_x$)Fe$_2$As$_2$ compounds with nominal x=0 to 0.40
by solid state reaction method. We found that the overall behavior
of (Sr$_{1-x}$K$_x$)Fe$_2$As$_2$ series is very similar to ReOFeAs
(Re=La, Ce, Pr, Nd, Sm). The observed anomaly at 210 K is
attributed to a spin-density-wave (SDW) instability. K-doping
suppresses this anomaly and induces superconductivity. This
strongly suggests that the competing orders are the common feature
for Fe$_2$As$_2$ tetrahedral layers being present in both
ZrCuSiAs- and ThCr$_2$Si$_2$-type structures. We performed the
Hall coefficient measurement, and confirmed that the dominant
carriers are hole-type with much higher carrier density compared
with that of single FeAs layer LaO$_{0.9}$F$_{0.1}$FeAs.

\begin{acknowledgments}
We acknowledge help from R. C. Yu and Q. M. Meng in experiments.
This work is supported by the National Science Foundation of
China, the Knowledge Innovation Project of the Chinese Academy of
Sciences, and the 973 project of the Ministry of Science and
Technology of China.

\end{acknowledgments}


\end{document}